\documentclass[prl,a4paper,twocolumn,showpacs,superscriptaddress]{revtex4}
\pdfoutput=1
\usepackage{amsmath}
\usepackage{amsfonts}
\usepackage{amssymb}
\usepackage{bm}
\usepackage{graphicx}
\bmdefine\bomega{\omega} \bmdefine\bOmega{\Omega}
\bmdefine\bnabla{\nabla} \bmdefine\bkappa{\kappa}
\bmdefine\bphi{\phi}

\begin{document}

\title{Precessing Vortex Motion and Instability in a Rotating Column
of Superfluid $^3$He-B}

\author{R.~H\"anninen} \affiliation{Low Temperature Laboratory, Helsinki University of
Technology, P.O.Box 5300, 02015 HUT, Finland }

\author{V.B.~Eltsov} \affiliation{Low Temperature Laboratory, Helsinki University of
Technology, P.O.Box 5300, 02015 HUT, Finland }
\affiliation{Kapitza Institute for Physical Problems, Kosygina 2,
119334 Moscow, Russia}

\author{A.P.~Finne} \affiliation{Low Temperature Laboratory, Helsinki University of
Technology, P.O.Box 5300, 02015 HUT, Finland }

\author{R. de Graaf} \affiliation{Low Temperature Laboratory, Helsinki University of
Technology, P.O.Box 5300, 02015 HUT, Finland }

\author{J.~Kopu} \affiliation{Low Temperature Laboratory, Helsinki University of
Technology, P.O.Box 5300, 02015 HUT, Finland }

\author{M.~Krusius} 
\affiliation{Low Temperature Laboratory, Helsinki University of
Technology, P.O.Box 5300, 02015 HUT, Finland }

\author{R.E. Solntsev} \affiliation{Low Temperature Laboratory, Helsinki University of
Technology, P.O.Box 5300, 02015 HUT, Finland }

\date{\today}

\begin{abstract}

The flow of quantized vortex lines in superfluid $^3$He-B is
laminar at high temperatures, but below $0.6\,T_{\rm c}$
turbulence becomes possible, owing to the rapidly decreasing
mutual friction damping. In the turbulent regime a vortex evolving
in applied flow may become unstable, create new vortices, and
start turbulence. We monitor this single-vortex instability with
NMR techniques in a rotating cylinder. Close to the onset
temperature of turbulence, an oscillating component in NMR
absorption has been observed, while the instability generates new
vortices at a low rate $\sim 1\,$vortex/s, before turbulence sets
in. By comparison to numerical calculations, we associate the
oscillations with spiral vortex motion, when evolving vortices
expand to rectilinear lines.

\end{abstract}

\pacs{67.57.Fg, 47.32.-y, 67.40.Vs} \maketitle

In superfluid dynamics a longstanding goal has been to account for
the formation of all vortices. A quantized vortex line is a
topologically stable structure of a coherent order parameter
field, with fixed circulation of superflow. What is then the
mechanism by which sudden burst-like turbulent vortex
proliferation is started from one single vortex loop which is
evolving in applied flow? The common view holds that turbulence,
once started, is sustained via loop formation and reconnections,
as seen in well-developed thermal counterflow turbulence of
superfluid $^4$He-II \cite{Vinen}. However, as emphasized by
Schwarz \cite{Schwarz}, to start turbulence, for instance in
linear flow along a circular pipe, vortices have to be pinned in
``vortex-mill'' configurations which continuously inject new
vortex loops downstream in the applied flow. Nevertheless, a
turbulent burst of vortex formation is observed to evolve from a
single seed vortex loop in rotating $^3$He-B, with no permanently
pinned vortices \cite{NeutronInjection}.

A recent explanation \cite{Precursor} of this controversy
concludes that a vortex evolving in applied flow may become
unstable while interacting and reconnecting with the container
wall. In $^3$He-B this \emph{single-vortex instability} becomes
possible with decreasing temperature below an onset temperature
$T_{\rm on}$, where mutual friction damping has dropped to
sufficiently low value: $\alpha (T) \lesssim 1$. In the rotating
cylinder the instability leads to a sudden transition to the
equilibrium vortex state. This process is most conveniently
studied in the onset temperature regime $T \sim T_{\rm on}$, by
monitoring the number of vortex lines as a function of time at
constant rotation velocity, as a response to the injection of a
seed vortex \cite{de Graaf:2007}. The events following the
instability have been reviewed in Refs.~\cite{ROP,PLTP} for the
case of a long rotating column. Here the end point of an evolving
seed vortex describes a spiral trajectory along the cylinder wall
while it expands towards its stable state as rectilinear vortex
line. In the onset region $T \sim T_{\rm on}$, the spiral vortex
motion is sometimes observed to give rise to an oscillating NMR
absorption signal which is examined in this report.

\begin{figure}[t]
\begin{center}
\centerline{\includegraphics[width=1\linewidth]{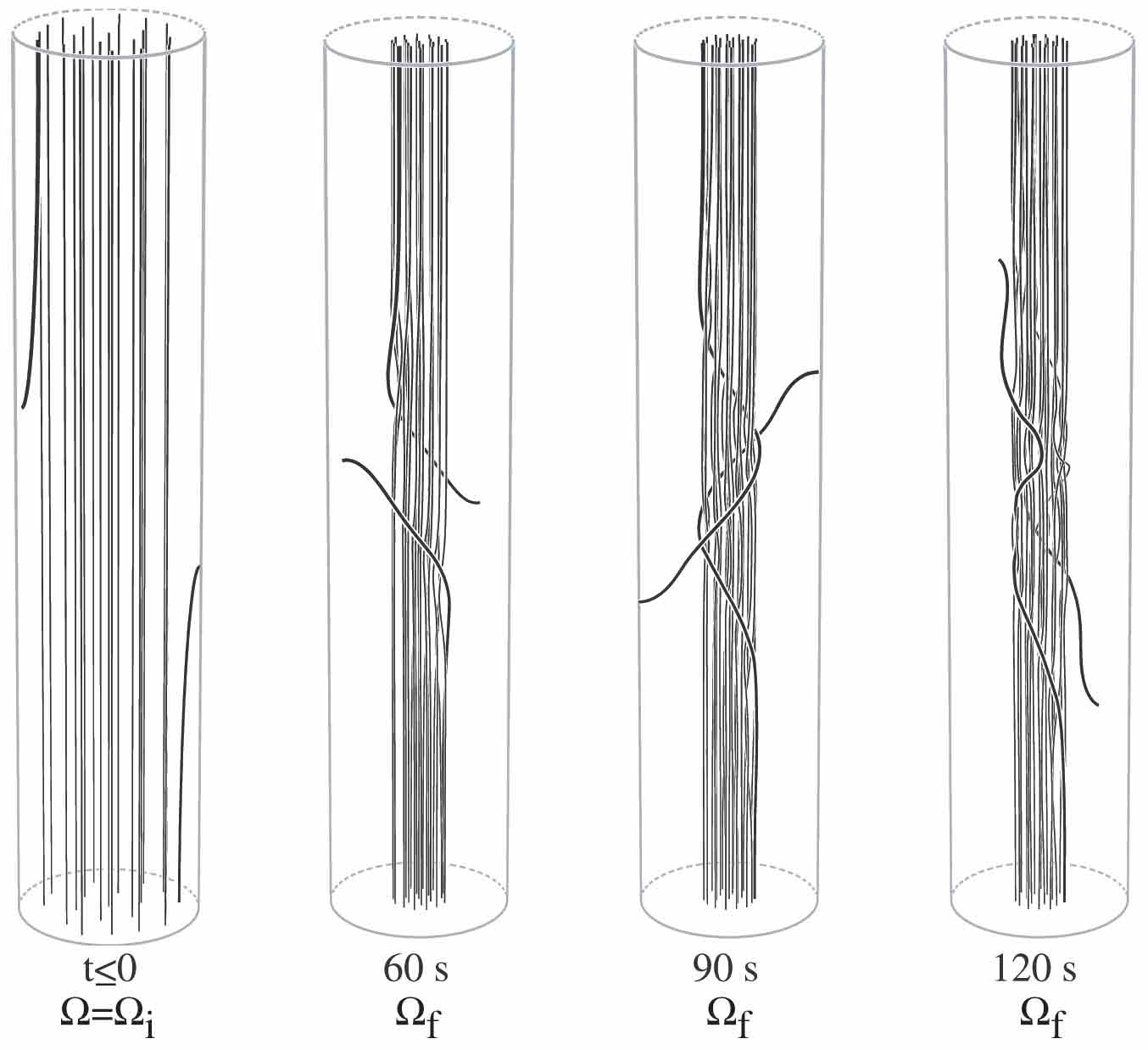}}
\caption{Numerical calculation of vortex evolution in a rotating
cylinder of $^3$He-B: Rotation is suddenly increased at $t=0$ from
$\Omega_{\rm i} \approx 0.03\,$rad/s to $\Omega_{\rm f} =
0.2\,$rad/s. There are 22 vortices in this sample, of which two in
the outermost ring (lying opposite to each other) have been
initially bent to the cylindrical wall, to break cylindrical
symmetry. In the later snapshots at $\Omega_{\rm f}$, the two
short vortices expand towards the top and bottom end plates of the
cylinder, while other vortices have contracted to a central vortex
cluster. Parameters: $R=3\,$mm, $L=30\,$mm, $P = 29.0\,$bar, and
$T=0.4\,T_{\rm c}$ (which corresponds to $\alpha = 0.18$ and
$\alpha^{\prime} = 0.16$ \cite{Bevan}).}
\label{EquilVorStateInject}
\end{center}
\vspace{-12mm}
\end{figure}

\begin{figure}[t]
\begin{center}
\includegraphics[width=0.45\linewidth]{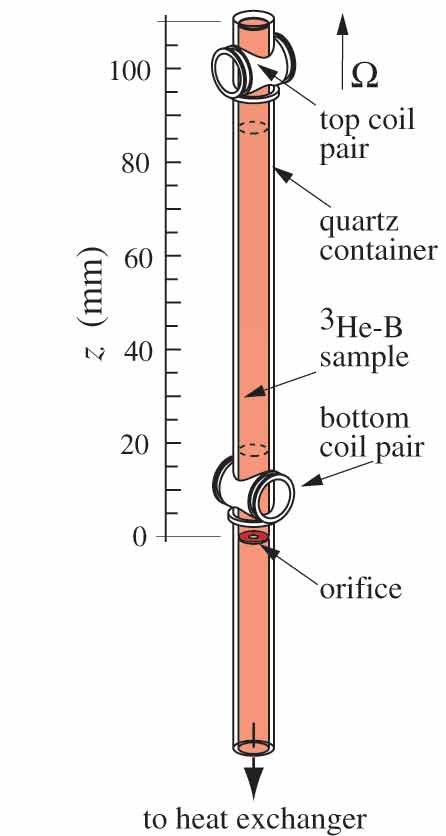}
\caption{Sample cylinder with detector coils. } \label{Sample}
\end{center}
\vspace{-6mm}
\end{figure}

{\bf Numerical illustrations:} To visualize the motion of evolving
vortices in rotating flow, a calculation is presented in
Fig.~\ref{EquilVorStateInject} using the vortex filament method
described in Ref.~\cite{de Graaf:2007}. The initial configuration
in Fig.~\ref{EquilVorStateInject}, with two curved vortices which
connect at one end to the cylindrical side wall, mimics the
equilibrium vortex state in a real rotating experiment.  Depending
on the rotation velocity $\Omega_{\rm i}$ and the residual
misalignment between the rotation and the sample cylinder axes ($
\lesssim 1^{\circ}$ in the setup of Fig.~\ref{Sample}), a certain
fraction of the peripheral vortices ends on the cylindrical wall
\cite{misalignment}.

\begin{figure}[t]
\begin{center}
\centerline{\includegraphics[width=0.8\linewidth]{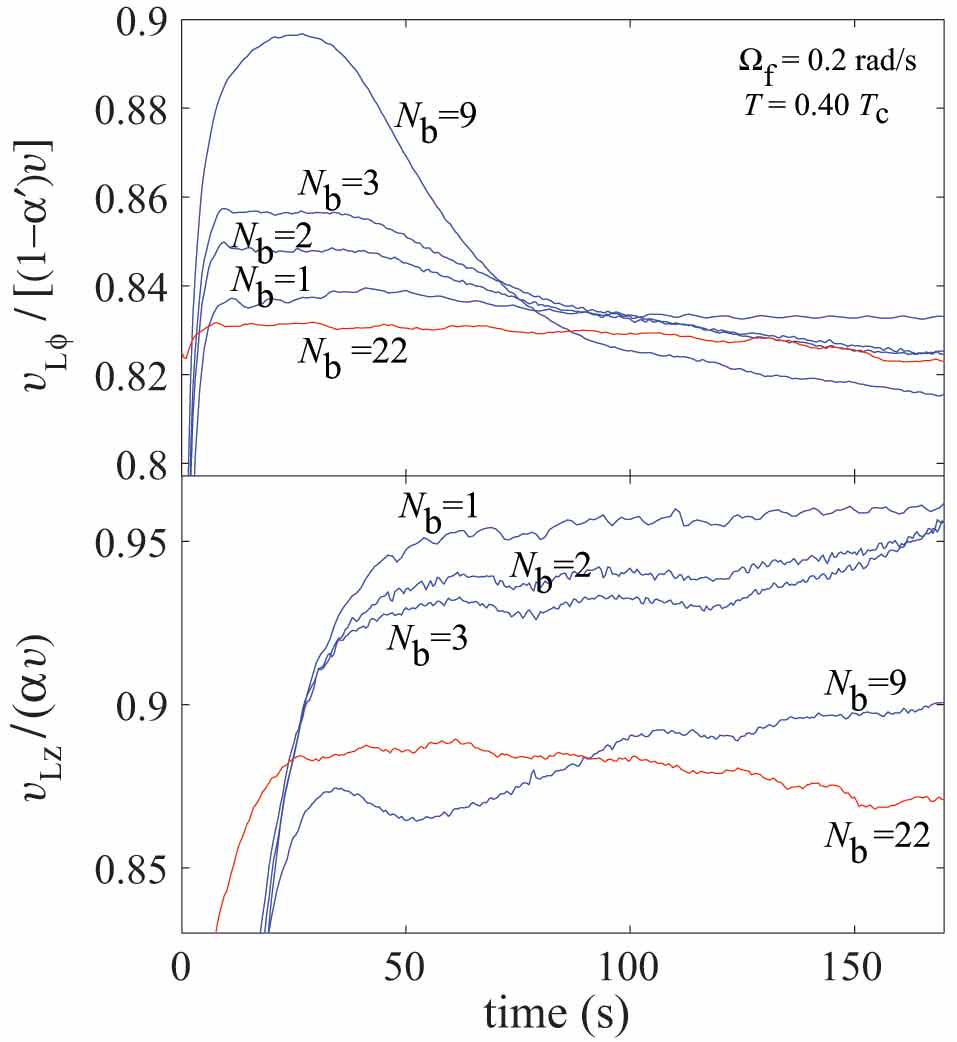}}
\caption{Comparison of azimuthal $v_{{\rm L}{\phi}}$ (top) and
longitudinal $v_{{\rm L}z}$ (bottom) velocities when different
numbers of $2N_{\rm b}$ vortices expand in the setup of
Fig.~\protect\ref{EquilVorStateInject}. With 22 vortices in total,
1, 2, 3, or 9 vortices of equal initial length $0.4\,L$ have been
bent both on the top and bottom to the cylindrical wall, to mimic
a tilted cylinder, as shown in
Fig.~\protect\ref{EquilVorStateInject}. The spiralling motion both
upwards and downwards along the rotating column is calculated as a
function of time. The average velocity of the vortex ends on the
cylindrical wall in the upward and downward moving bundles is
plotted during the time needed to reach the respective end plate.
The actual velocity is the number on the vertical scale times
$\alpha v(\Omega_{\rm f},R,N)$ or
$(1-\alpha^{\prime})v(\Omega_{\rm f},R,N)$, where $N = 22-N_{\rm
b}$. The case $N_{\rm b} = 22$ is different: here all 22 vortices
are initially bent at one cylinder end to the wall and, after the
rotation increase to $\Omega_{\rm f}$, a vortex front starts
expanding towards the other vortex-free end of the cylinder. This
situation is examined further in Fig.~\ref{PrecessVorFront}.}
\label{SpiralVorVelocities}
\end{center}
\vspace{-8mm}
\end{figure}

In Fig.~\ref{EquilVorStateInject}, rotation is increased in
step-like manner at $t=0$ from the equilibrium vortex state at
$\Omega_{\rm i} \neq 0$ to a final stable value $\Omega_{\rm f} $.
The $N= 20$ rectilinear vortices are thereby compressed to a
central cluster with an areal density $n_{\rm v} = 2 \Omega_{\rm
f} /\kappa$ by the surrounding counterflow (cf) at the velocity
$\bm{v} = \bm{v}_{\rm n} - \bm{v}_{\rm s}$, the difference between
the velocities of the normal and superfluid fractions. The highly
viscous normal component we consider to be in solid-body rotation:
$\bm{v}_{\rm n} = \bm{\Omega} \times \bm{r}$. Outside the
compressed cluster the cf velocity is then given by
\begin{equation}
v(\Omega_{\rm f},r,N) = v_{\rm n} - v_{\rm s} = \Omega_{\rm f} r -
\frac{\kappa N} {2\pi r} \, , \label{cf}
\end{equation}
where $\kappa = h/(2m_3)$ is the quantum of circulation. The two
short vortices now expand in this cf in spiral motion towards the
top and bottom end plates, respectively. The velocity ${\bf
v}_{\rm L}$ of a vortex line element with tangent
$\hat{\mathbf{s}}$ is obtained from the equation of motion
\cite{Donnelly}
\begin{equation}
\bm{v}_{\rm L}= \bm{v}_{\rm s} +\alpha \hat{\mathbf{s}} \times
(\bm{v}_{\rm n}-\bm{v}_{\rm s}) -\alpha' \hat{\mathbf{s}} \times
[\hat{\mathbf{s}}\times(\bm{v}_{\rm n}-\bm{v}_{\rm s} )]\,
,\label{vl}
\end{equation}
where $\alpha$ is the dissipative and $\alpha^{\prime}$ the
reactive mutual friction coefficient. Since the vortex end is
perpendicular to the cylindrical wall, it has from Eq.~(\ref{vl})
a longitudinal velocity $v_{{\rm L}z} = \alpha v(\Omega_{\rm
f},R,N)$ and an azimuthal component $v_{{\rm L}{\phi}} =
(1-\alpha^{\prime}) v(\Omega_{\rm f},R,N)$. In the rotating
coordinate system, $v_{{\rm L}{\phi}}$ points in the opposite
direction from the rotation of the container. The resulting end
point motion gives us a simple picture of the spiral trajectories
along the cylindrical wall, although other parts of the vortex
also contribute to its motion, in particular its curvature where
it connects to the cylindrical wall. During their expansion the
two evolving vortices are wound around the central vortex cluster
as a helix whose pitch depends on the cf velocity $v(\Omega_{\rm
f},R,N)$. Ultimately, the helix will unwind, when the vortex ends
slip along the flat end plates of the cylinder, so that the final
state is composed of only rectilinear vortex lines. However, since
the cf velocity is close to zero at the border of the cluster, the
unwinding is a slow process.

The calculated velocities of the vortex ends in
Fig.~\ref{EquilVorStateInject} are $v_{{\rm L}z} \approx 0.84 \,
\alpha \Omega R \approx 0.96 \, \alpha \, v(\Omega_{\rm f},R,N)$
and $v_{{\rm L}{\phi}} \approx 0.73 (1-\alpha^{\prime}) \Omega R
\approx 0.83(1-\alpha^{\prime}) \, v(\Omega_{\rm f},R,N)$. The
wave length of the spiral trajectory is thus $\lambda = $ $2\pi R
\, v_{{\rm L}z}/v_{{\rm L}{\phi}} \approx 5\,$mm and the period $p
= 2\pi R/v_{{\rm L}{\phi}} \approx \,50\,$s. In
Fig.~\ref{SpiralVorVelocities} the spiral vortex motion is
analyzed further for the same experimental setup as in
Fig.~\ref{EquilVorStateInject}, but when several vortices are
expanding simultaneously. The initial starting state for these
calculations is one where a specified number ($N_{\rm b}$) of
nearest-neighbor vortices is bent to the cylindrical wall at the
same distance of $0.4\,L$ from the end plate both at the bottom
and top of the cylinder, similar to the configuration in
Fig.~\ref{EquilVorStateInject} (at $t=0$). When rotation is
increased from $\Omega_{\rm i}$ to $\Omega_{\rm f}$, one set of
$N_{\rm b}$ curved vortices starts expanding towards the top and
one towards the bottom end plate. The axial (bottom panel) and
azimuthal velocities (top panel) have been plotted as a function
of time, for different numbers of $2N_{\rm b}$ vortices in spiral
expansion.

Two observations can be made from Fig.~\ref{SpiralVorVelocities}.
The vortices, which expand in one direction and initially start
off in close proximity of each other, remain a close bundle during
their spiral motion.  Vortex bundling is a general phenomenon
which gives rise to the formation of larger eddies, similar to the
eddies in classical viscous fluid motion. Here the bundles do not
disperse, but are well preserved especially during the later part
of the expansion in a state of steady propagation, since all
$N_{\rm b}$ vortices travel roughly at the same speed both axially
and azimuthally along the cylinder. Thus the spread in velocities
among the different vortices within the bundle and also the
changes in the velocities as a function of time remain relatively
small. The second note about Fig.~\ref{SpiralVorVelocities} is
that both $v_{{\rm L}z}$ and $v_{{\rm L}{\phi}}$ are rather
insensitive to the number of vortices spiralling in the bundle.
This is because the total number of vortices $N = 22$ corresponds
here to a very low equilibrium rotation velocity $\sim
0.03\,$rad/s compared to the actual rotation velocity of
0.2\,rad/s. For better comparison with measurements the final
rotation velocity $\Omega_{\rm f}$ should be increased even more,
which calls for time-consuming calculations.

\begin{figure}[t]
\begin{center}
\centerline{\includegraphics[width=0.85\linewidth]{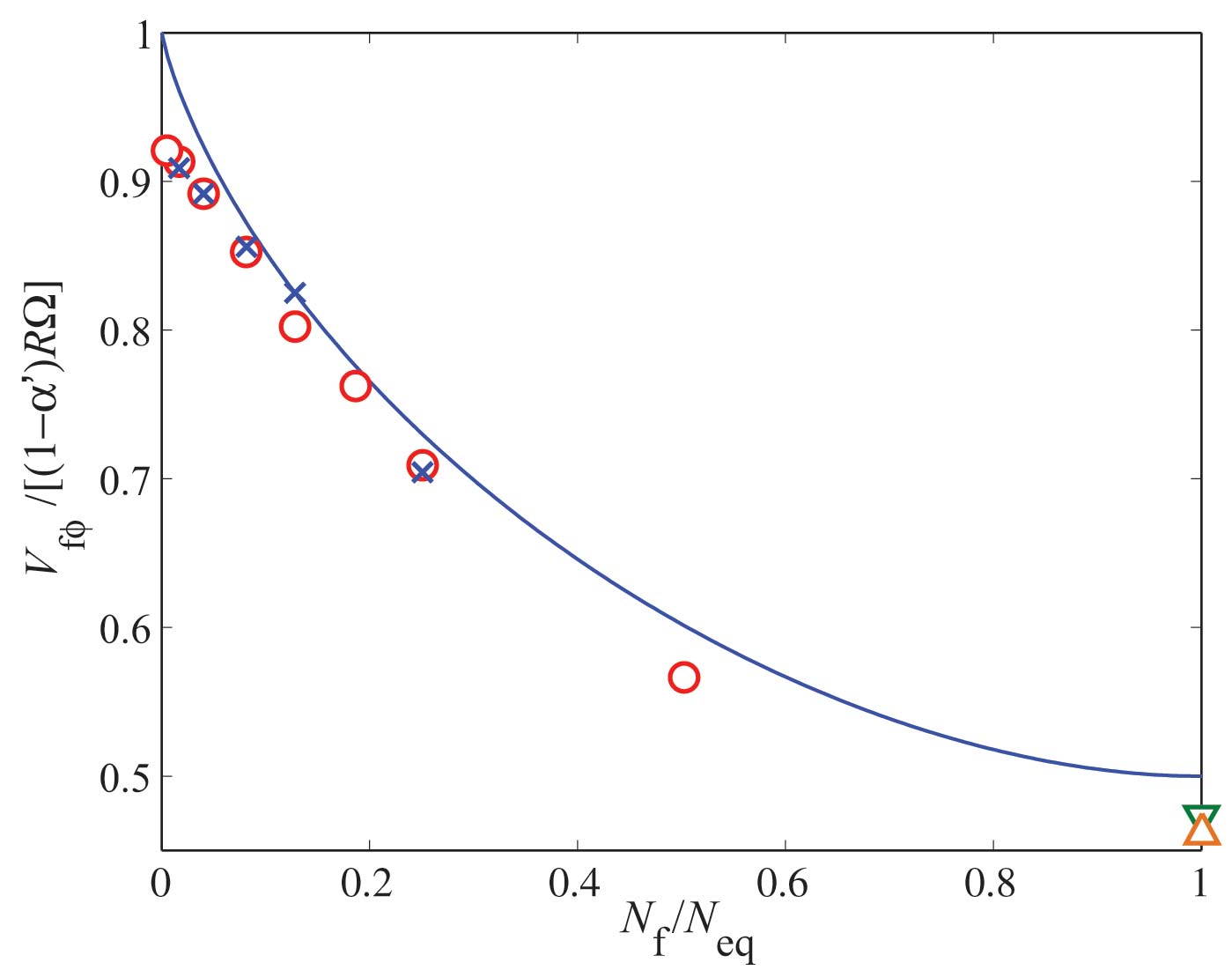}}
\caption{Calculated azimuthal front velocity $V_{{\rm f}\phi}$ (in
the rotating coordinate system), as a function of the number of
vortices $N_{\rm f}$ which compose the front. No prior vortex
cluster exists in the center of the cylinder. The zero temperature
limit of Eq.~\ref{vm} {\it (solid curve)} is compared to
calculations at finite temperatures {\it (data points)}. The
calculations with $N_{\rm f} < N_{\rm eq}$ are for $R = 3\,$mm, $L
= 30\,$mm, $\Omega = 0.2\,$rad/s, where $N_{\rm eq} \approx 170\,$
vortices: ($\circ$) $\alpha = 0.1$, $\alpha^{\prime} =0$;
($\times$) $\alpha = 0.18$, $\alpha^{\prime} = 0.16$, $T \approx
0.4\,T_{\rm c}$. The two calculations with $N_{\rm f} \approx
N_{\rm eq}$ are for $R = 1.5\,$mm, $L = 40\,$mm, $\Omega =
1\,$rad/s, where $N_{\rm eq} \approx 210\,$ vortices:
($\bigtriangledown$) $\alpha = 0.040$, $\alpha^{\prime} = 0.030$,
$T \approx 0.3\,T_{\rm c}$; ($\bigtriangleup$) $\alpha = 0.18$,
$\alpha^{\prime} = 0.16$, $T \approx 0.4\,T_{\rm c}$. On the
vertical scale the normalized velocity $V_{{\rm f}\phi}/[(1-
\alpha^{\prime}) \Omega R]$ is plotted. $V_{{\rm f}\phi}$ is taken
as the front velocity when a stable value has been reached, which
generally happens when the front has propagated 2/3 of the length
of the cylinder. } \label{PrecessVorFront}
\end{center}
\vspace{-8mm}
\end{figure}

\begin{figure}[t]
\begin{center}
\includegraphics[width=1\linewidth]{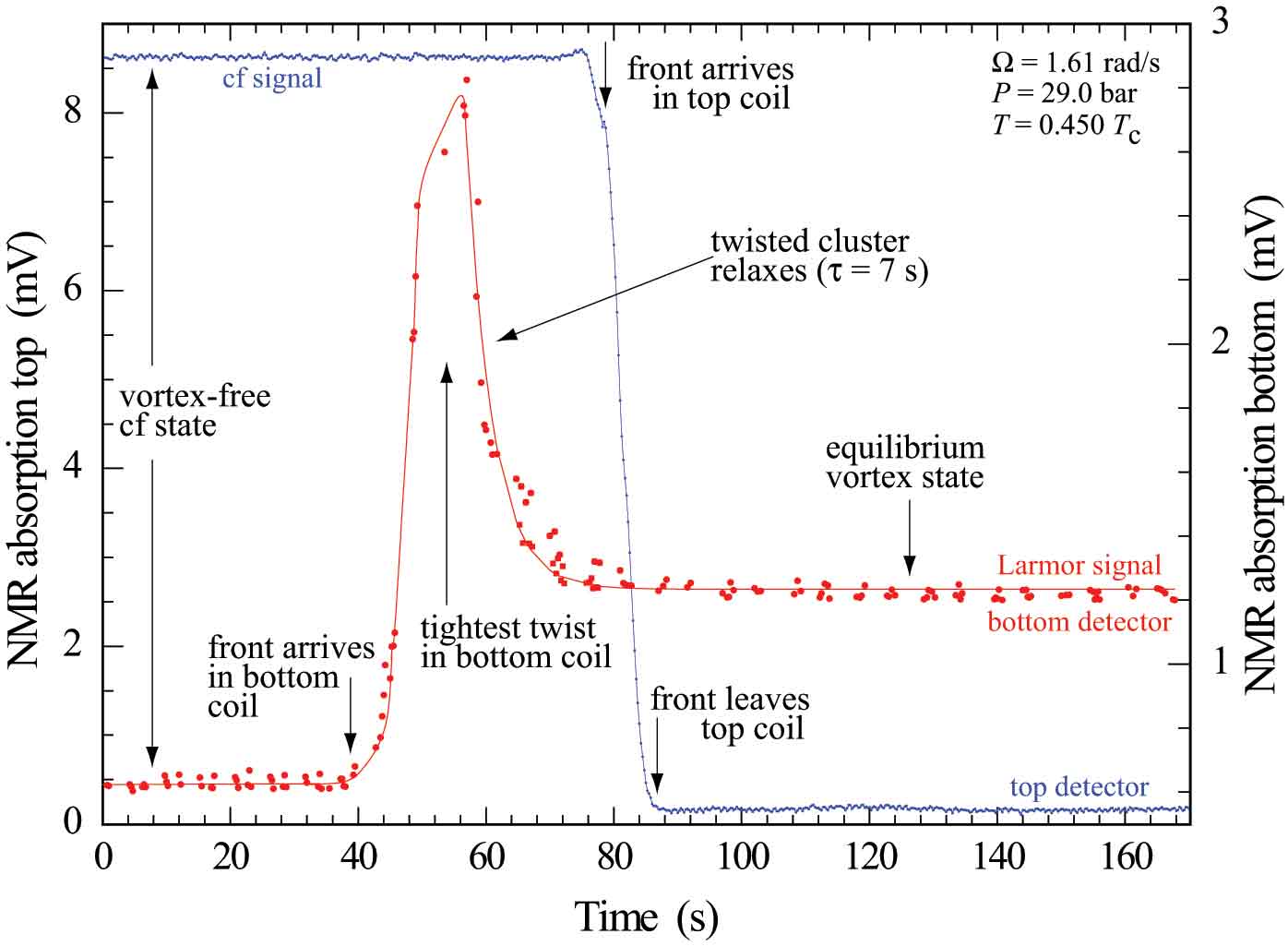}
\caption{NMR response during a rapid transition from vortex-free
flow to the equilibrium vortex state at constant externally
controlled conditions. The top and bottom detectors monitor the
NMR absorption at different values of constant NMR field. The cf
peak height is recorded with the top detector, while the bottom
detector monitors the peak close to the Larmor edge. Time $t=0$
marks the moment when $\Omega_{\rm f} = 1.61\,$rad/s is reached.
The time delay between the sudden rise in the bottom response (at
$ t = 40\,$s) and the start of the rapid collapse in the top (at $
t = 78\,$s) is caused by the longitudinal motion of the two vortex
fronts along the column, controlled by $\alpha = 0.33$
\cite{Bevan}. } \label{FastCF_PeakResponse}
\end{center}
\vspace{-8mm}
\end{figure}

A special case is that where all existing vortices are released
simultaneously from one end of the rotating cylinder and expand as
a vortex front towards the vortex-free end of the cylinder
(\textit{i.e.} where no prior central vortex cluster exists). In
Fig.~\ref{SpiralVorVelocities} this case is the example with
$N_{\rm b} = 22$. Such a calculation models the experimental
situation after a turbulent burst, like in
Fig.~\ref{FastCF_PeakResponse} (although there the burst occurs
higher up in the column, which then starts both an upward and
downward moving front). At temperatures below about $0.45\, T_{\rm
c}$ the front, once it is formed from $N_{\rm f}$ vortices,
maintains its constant narrow width during the spiralling
propagation along the column \cite{PLTP,ROP}. The longitudinal
propagation velocity $V_{\rm fz}$ of the front has been measured
and analyzed in Ref.~[\onlinecite{FlightTime}]. The azimuthal
velocity $V_{{\rm f}{\phi}}$ (in laboratory coordinates) is
obtained in the continuum model at zero temperature (where
friction vanishes) from the equation
\begin{equation}
V_{{\rm f}{\phi}}= \frac{\kappa N_{\rm f}} {2\pi R} \; \frac{
1/2\,\ln{(N_{\rm eq} / N_{\rm f})} + 1/4} {1- N_{\rm f} /(2 N_{\rm
eq})}\, , \label{vm}
\end{equation}
where the approximate continuum value for the number of vortices
in the equilibrium state is $N_{\rm eq} = \pi R^2
(2\Omega/\kappa)$. When $N_{\rm f} \rightarrow N_{\rm eq}$, this
expression gives $V_{{\rm f}{\phi}}= {\frac{1}{2}}\,\Omega R$.
This limit was discussed in Ref.~\cite{TwistedPropagation}. In
Fig.~\ref{PrecessVorFront} we compare Eq.~(\ref{vm}) to numerical
calculations at finite temperatures and finite friction for a few
examples. The results for the normalized azimuthal front velocity
$V_{{\rm f}\phi}/[(1- \alpha^{\prime}) \Omega R]$ drop below the
estimate from Eq.~(\ref{vm}), but by less than 10\,\%. Thus also
at finite friction $V_{{\rm f}{\phi}} \approx \frac{1}{2}
(1-\alpha^{\prime}) \Omega R$ is a good approximation, which
applies when the number of vortices in the front is large ($N
\lesssim N_{\rm eq}$).

\begin{figure}[t]
\begin{center}
\centerline{\includegraphics[width=1\linewidth]{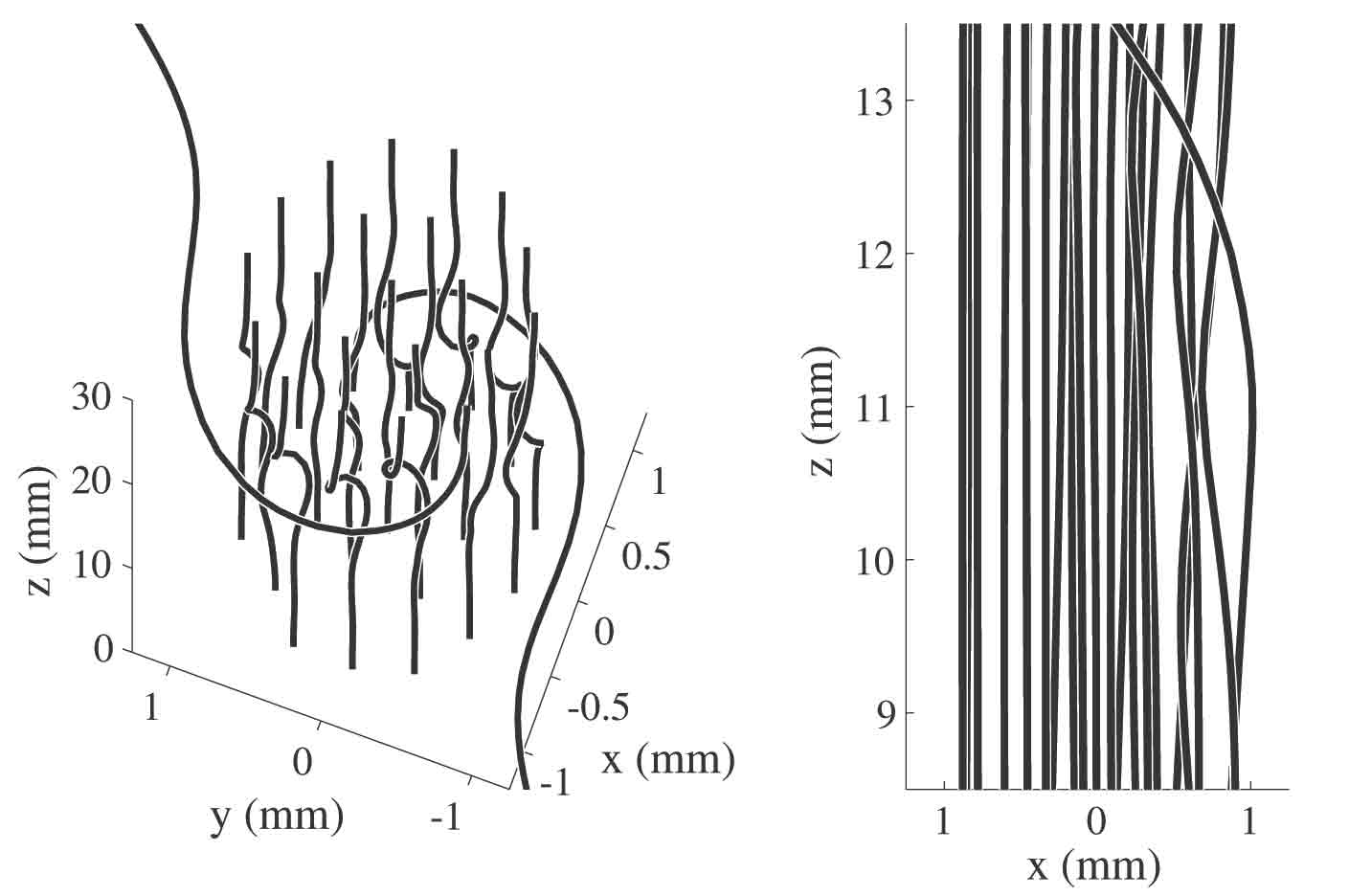}}
\caption{{\it (Left)} Rotating vortex cluster from
Fig.~\ref{EquilVorStateInject} at $t=60\,$s, viewed at an angle
deviating by $3^{\circ}$ from vertical. Only the central part of
the sample cross section in the $x$ -- $y$ plane is shown. Helical
Kelvin waves are induced on the vortex lines in the cluster by the
two vortices spiralling outside the cluster. In the early stage of
the expansion, the waves start at roughly half-height of the
cylinder, \textit{i.e.} from the points where the two short
vortices merge with the cluster (see
Fig.~\protect\ref{EquilVorStateInject}). {\it (Right)} Side view
through the vortex cluster ($t=60\,$s). A vortex spiralling around
the cluster induces radial displacements on the outermost vortices
within the cluster. The displacements start longitudinally
propagating Kelvin waves on these vortices. }
\label{PrecessVorMotion}
\end{center}
\vspace{-8mm}
\end{figure}

While spiralling around the cluster in
Fig.~\ref{EquilVorStateInject}, the two curved vortices may
reconnect with each other or with the outermost lines in the
cluster. These reconnections do not increase the vortex number. As
can be seen from  Figs.~\ref{EquilVorStateInject} and
\ref{PrecessVorMotion}, a spiralling vortex distorts the cluster
by inducing propagating helical Kelvin waves on the vortex lines,
which results in oscillations of the cluster around its
equilibrium position \cite{movies}. The amplitude of the waves is
comparable to the inter-vortex distance in the cluster, $\sim
1/\sqrt{n_{\rm v}}$, and initially the wavelength and period of
the excited Kelvin waves have similar values as the precessing
vortex motion outside the cluster.

At high friction ($\alpha \geq 0.18$) the number of vortices
remains strictly constant in Figs.~\ref{EquilVorStateInject} --
\ref{PrecessVorMotion}. At low friction ($\alpha \leq 0.1$) the
calculations on the precessing vortex front in
Fig.~\ref{PrecessVorFront} may display some increase in vortex
number, when $N < N_{\rm eq}$. Figs.~\ref{EquilVorStateInject} --
\ref{SpiralVorVelocities} thus illustrate the situation above
$T_{\rm on}$ where no increase in vortex number occurs.
Experimentally $T_{\rm on}$ depends on the applied flow velocity
($\sim [\Omega_{\rm f} - \Omega_{\rm i}] R$) and on the number
($N_{\rm f}$) and initial configuration of evolving vortices
\cite{Remnants,de Graaf:2007,PLTP}. In most cases $T_{\rm on}$ is
found to be higher than the temperature of the present
calculations, $0.40\, T_{\rm c}$. The reason for this difference
appears to be the high stability of rotating flow in a circular
cylinder, which according to our calculations \cite{de Graaf:2007}
appears to be a particular characteristic of a cylinder with a
circular cross section and ideal wall properties, \textit{i.e.}
without surface friction or pinning. The most likely section of an
evolving vortex to become unstable is the curved piece ending on
the cylindrical wall in the maximum flow $v(\Omega_{\rm f},R,N)$,
while the most stable parts are the sections which reside in low
cf close to the central vortex cluster. The extreme case are
rectilinear vortex lines in the cluster which are experimentally
found to remain stable even in sinusoidally modulated rotation at
temperatures $\gtrsim 0.3\,T_{\rm c}$.

{\bf Experiment:} The sample is cooled in a rotating nuclear
demagnetization cryostat at 29\,bar liquid pressure within a fuzed
quartz cylinder, with radius $R=3\,$mm and length $L=110\,$mm
(Fig.~\ref{Sample}). The number of vortices is measured
non-invasively simultaneously at both ends of the long cylinder,
by monitoring the NMR absorption line shape. Both NMR detectors
consist of circular split-pair coils, with a separation of 9\,mm
between the coil halves and with their common axis aligned
transverse to the cylinder. The coils are wound from
superconducting wire and are part of a LC resonator with a Q value
of $\sim 1\cdot 10^4$, coupled to a GaAs MESFET preamplifier
located within the cryostat at 4.2\,K temperature. Both NMR
spectrometers operate independently at constant frequency in an
axially oriented polarization magnetic field with a linear sweep.
The NMR setup is described in more detail in
Refs.~\cite{NeutronInjection,ExpSetup}

With vortex-free cf, the NMR absorption line shape includes a
prominent cf peak which is shifted far from the Larmor frequency,
while in the equilibrium vortex state the absorption drops to zero
at the location of the cf peak \cite{Kopu}. The height of the cf
peak increases with the magnitude of the azimuthal cf velocity
$v(\Omega,R,N)$ and thus decreases with the number of vortices $N$
in the central cluster. The correspondence between peak height and
vortex number $N$ can be obtained from calculations of the order
parameter texture or from measurements. The measurements which we
discuss below use remanent vortices \cite{Remnants} as the seeds
which evolve in the applied flow after the rotation increase to
$\Omega_{\rm f}$. Such a measurement is performed at constant
temperature, by first decelerating rotation to zero from a state
with a large number of vortices. Zero rotation is then maintained
for a period $\Delta t$, to allow vortices to annihilate, with the
exception of a few remaining dynamic remnants \cite{Remnants}. The
final step is to increase $\Omega$ at a fixed rate (typically $d
\Omega/dt \leq 0.02\,$rad/s$^2$) to $\Omega_{\rm f}$, where it is
kept constant while the buildup in the number of vortex lines
$N(t)$ in the central cluster is recorded as a function of time.

{\bf Single vortex instability:} Well above the onset temperature,
the cf peak height settles at $\Omega_{\rm f}$ to a stable value
and remains constant. Well below onset the cf peak collapses to
zero when the equilibrium vortex state is formed. Around the onset
temperature $T \approx T_{\rm on}$, both types of behavior occur
randomly. If the equilibrium vortex state is here formed, then in
perhaps one third of the cases the collapse of the cf peak can be
preceded by an initial slow decrease in peak height. The slow
reduction in peak height corresponds to slow generation of new
vortices. It is the signature of the single-vortex instability, as
described and analyzed in Ref.~\cite{de Graaf:2007}. It can only
be observed in the onset regime $T \sim T_{\rm on}$; at lower
temperatures the instability proceeds too rapidly to be monitored
with our techniques. The slow vortex generation is terminated in a
turbulent burst which takes place as a localized event in some
short section (of length $\sim R$) of the long sample column. In
this burst enough vortices are created to start the evolution
towards the equilibrium vortex state. From the site of the burst
the vortices propagate in spiral motion as a front both up and
down along the cylinder, leaving behind a vortex bundle composed
of helically twisted vortices \cite{ROP,PLTP}.

\begin{figure}[t]
\begin{center}
\includegraphics[width=1\linewidth]{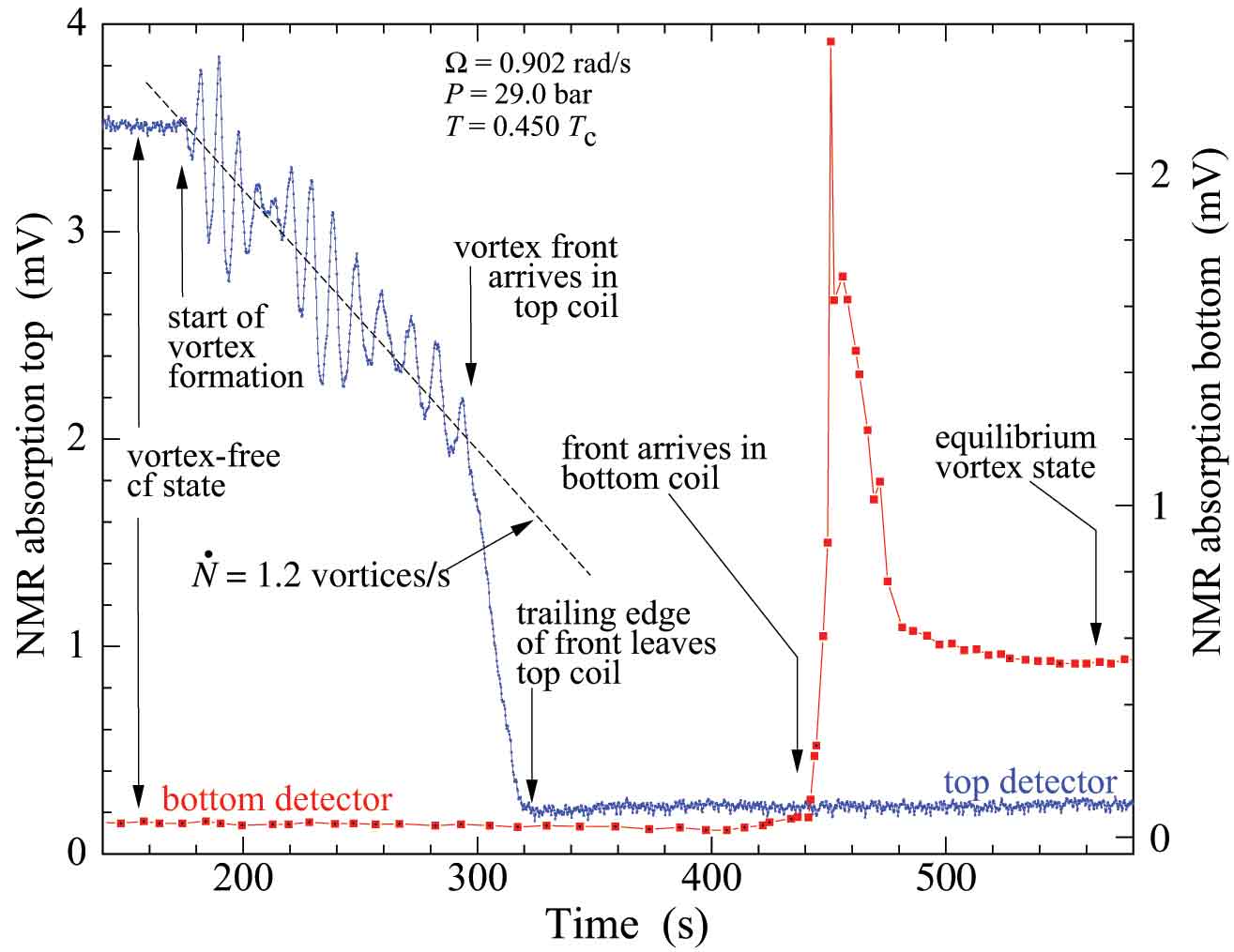}
\caption{Precursory vortex formation which terminates in a sudden
turbulent burst. This is a repetition of the measurement in
Fig.~\ref{FastCF_PeakResponse}, but at a lower rotation velocity
$\Omega_{\rm f} = 0.90\,$rad/s. The difference is the prolonged
slow vortex generation via the single-vortex instability, denoted
by the dashed line. Superimposed comes the large-amplitude
quasi-coherent oscillation in the cf peak height at a frequency
which is related to the azimuthal vortex motion. The latter is
controlled by $1-\alpha^{\prime}= 0.76$ \cite{Bevan}. }
\label{OscCF_Peak}
\end{center}
\vspace{-8mm}
\end{figure}

The NMR absorption responses of rapid and slow transitions to the
equilibrium vortex state are compared in
Figs.~\ref{FastCF_PeakResponse} and \ref{OscCF_Peak},
respectively. These were recorded in two consecutive measuring
runs at the same temperature using the same measuring procedure.
The order of the two measurements was such that
Fig.~\ref{OscCF_Peak} was measured first and
Fig.~\ref{FastCF_PeakResponse} next. The responses prove to be
different owing to the different value of the rotation velocity
$\Omega_{\rm f}$, which in turn is caused by the stochastic nature
of the remanent vortex injection process: in the former case the
waiting period at zero rotation prior to the measurements was
$\Delta t = 40\,$min, while in the latter case it was 22\,min.
Thus in Fig.~\ref{FastCF_PeakResponse} the remanent vortices are
expected to be fewer and smaller loops. In both cases the sample
was then accelerated to rotation at $\dot{\Omega} =
0.004\,$rad/s$^2$. The acceleration was interrupted occasionally,
to check at constant rotation whether vortex-free rotation
continued to persist over long periods in time. In
Fig.~\ref{FastCF_PeakResponse} (Fig.~\ref{OscCF_Peak}) rotation
could be increased until 1.61\,rad/s (0.90\,rad/s), before any
indication of vortex formation was noticed. Time $t=0$ coincides
with the moment when $\Omega_{\rm f} = 1.61\,$rad/s (0.90\,rad/s)
was reached. After 40\,s (170\,s) the first response from vortices
is observed in the bottom (top) spectrometer. This should be
compared with the time it takes for a small vortex loop to expand
in vortex-free flow at $\Omega_{\rm f}$ from one end of the
cylinder to the other, to become a rectilinear line, $\approx
L/(\alpha \Omega_{\rm f} R) = 70\,$s (120\,s). In
Fig.~\ref{OscCF_Peak} the start of vortex formation is thus
unusual because of the long 170\,s delay, before the generation of
new vortices starts. The explanation for the long delay must be
associated with the properties of a small remanent vortex loop
with a radius close to the critical value of $\sim 4\,\mu$m needed
to overcome the barrier for spontaneous expansion \cite{Ruutu}.

{\bf Counterflow peak response :} In
Figs.~\ref{FastCF_PeakResponse} and \ref{OscCF_Peak} the peak
heights of two different absorption maxima of the NMR spectrum are
recorded as a function of time at constant externally controlled
conditions. These are: (1) The cf peak height, which initially in
vortex-free flow is at maximum and zero in the final equilibrium
vortex state. (2) The Larmor peak height, which initially is close
to zero, but in the final state different from zero. While the
Larmor peak response is qualitatively similar in the two examples,
the drop in the cf peak height happens differently.
Fig.~\ref{FastCF_PeakResponse} shows the generic response from a
rapid transition, while in Fig.~\ref{OscCF_Peak} the decay in the
cf peak height starts first with a more gentle linear decrease
(dashed line) where the number of vortices gradually increases
within the top detector, as generated by the single-vortex
instability. This continues for more than 100\,s, before the rapid
peak height decay starts. (Also in Fig.~\ref{FastCF_PeakResponse}
at higher $\Omega_{\rm f}$ a short vestige of slower peak height
reduction can be distinguished, but the rate is faster and it only
lasts for $\sim 4\,$s.) However most importantly, in
Fig.~\ref{OscCF_Peak} the slow decrease in peak height differs
from usual examples in that the cf peak height also displays well
resolved oscillations of large amplitude. The characterization of
these oscillations is the central issue in this report.

According to the measured calibration, the linear decrease in the
cf peak height, as marked by the dashed line in
Fig.~\ref{OscCF_Peak}, represents a vortex formation rate $\dot{N}
\approx 1.2\,$vortices/s. The later rapid collapse is caused by
the arrival of the vortex front \cite{FlightTime} to the bottom
end of the top coil and its subsequent travel through the coil. At
the arrival of the front, the central cluster within the top coil
contains $\sim 140$ vortices (to be compared with $N_{\rm eq}
\approx 840$ in the equilibrium vortex state). The front
incorporates approximately all the remaining vortices needed to
fill the sample with the equilibrium vortex state. The front,
composed of these additional vortices, travels in spiral motion
along the cylinder around the central cluster. Behind the front
the expanding vortices are wound in a helically twisted
configuration which later slowly relaxes
\cite{TwistedPropagation}. At that point the macroscopic cf, which
was generated by the rotation increase to $\Omega_{\rm f}$, is
completely removed.

{\bf Larmor peak response:} The second signal trace in
Figs.~\ref{FastCF_PeakResponse} and \ref{OscCF_Peak} is  recorded
by scanning the absorption maximum close to the Larmor edge with a
rapidly moving field sweep. Here the sudden sharp absorption
increase signals the arrival of the vortex front, as it passes
through the top edge of the bottom detector. The increasing
absorption is generated by the axially flowing supercurrent which
is produced by the helically twisted vortices behind the front
\cite{Kopu-2}. The subsequent exponential decay of this absorption
is caused by the unwinding of the twist while the vortex ends slip
along the top and bottom end plates of the sample cylinder and the
vortices are converted to rectilinear lines
\cite{TwistedPropagation}. Thus the Larmor absorption response
forms a transient peak, followed by a stable final absorption
level, which is representative of the equilibrium vortex state,
after the twist has relaxed \cite{ROP}.

In Fig.~\ref{OscCF_Peak} a delay time $t_{\rm d} \approx 144\,$s
separates the moments when the vortex fronts moving up and down
along the column pass through the bottom edge of the top detector
and the top edge of the bottom detector, respectively. This
distance is $82\,$mm. The longitudinal propagation velocity
$V_{{\rm f}z}$ of the front has been measured to be $V_{{\rm f}z}
/(\Omega R) \approx 0.28$ at $0.45\,T_{\rm c}$ \cite{FlightTime},
in a situation when there is no prior central vortex cluster.
Assuming that the propagation velocity of the front in the
presence of a central vortex cluster is $\approx v(\Omega_{\rm f},
R, N) \; V_{{\rm f}z} /(\Omega R)$, we estimate that, with a
vortex formation rate of $\dot{N} \approx 1.2\,$vortices/s, on an
average $N \approx 230\,$ vortices populate the central cluster at
the time when the front travels from the location of the turbulent
burst to the top edge of the bottom detector. The delay time of
144\,s can then be used to localize the site and the moment of the
turbulent burst. Here the burst proves to happen a couple of mm
below the top detector at $t \approx 290\,$s, or some five seconds
before the absorption response in the top detector starts to
collapse. In Fig.~\ref{FastCF_PeakResponse} at a higher flow
velocity all events occur faster: with a delay of 33\,s between
the first bottom and top responses, the burst happens at $t =
28\,$s some 19\,mm above the top edge of the bottom coil.

\begin{figure}[t]
\begin{center}
\centerline{\includegraphics[width=0.8\linewidth]{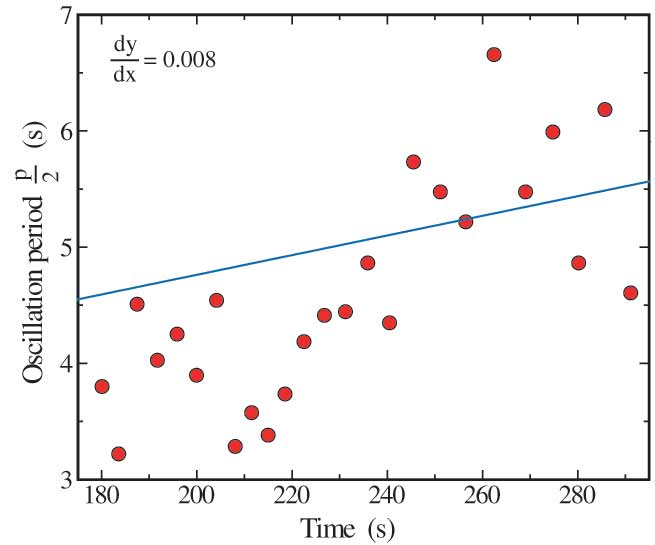}}
\caption{Time difference between the maxima and minima in the
oscillating cf peak height in Fig.~\protect\ref{OscCF_Peak}. The
line is obtained by inserting the appropriate values from
Fig.~\protect\ref{OscCF_Peak} in
Eq.~(\protect\ref{PrecessionFrequency}). } \label{OscPeriod}
\end{center}
\vspace{-8mm}
\end{figure}

{\bf Spiral vortex motion:} The generic signal from slow vortex
generation in the cf peak height is a relatively smooth decay,
while the many periods of well-resolved quasi-coherent oscillation
in Fig.~\ref{OscCF_Peak} are unusual. If both the top and bottom
detectors are tuned to monitor the cf peak height simultaneously,
then the two recordings of the oscillations look similar, but
neither the amplitudes or the frequencies are exactly identical.
Since the distribution and number of evolving vortices varies
along the column, this is expected. Obviously the oscillating
signal is driven by the spiral vortex motion, which is the only
source for precession at a frequency around 0.1\,Hz. In
Fig.~\ref{OscPeriod} we have extracted the difference in time
between the maxima and minima in the oscillation of
Fig.~\ref{OscCF_Peak}. This data set consists of 13 full periods.
The average period is 9.2\,s, but the period also appears to be
increasing with time. The spiral vortex motion is expected to have
the frequency
\begin{equation}
f_{\phi} = \varepsilon \, (1-\alpha^{\prime}) v(\Omega_{\rm f},
R,N) / (2\pi R)\;,\label{PrecessionFrequency}
\end{equation}
so that $\dot{f}_{\phi} \propto -\dot{N}$. We assume that the
azimuthal vortex velocity is of the form $v_{{\rm L}{\phi}} =
\varepsilon (1-\alpha^{\prime}) v(\Omega_{\rm f}, R, N)$, where
$\varepsilon < 1$ is a numerical factor, which in the case of
Fig.~\ref{SpiralVorVelocities} is about 0.8. The measuring setup
in Fig.~\ref{Sample} incorporates also an instrument factor: The
sensitivity of the detector coil pair is not constant across the
cross section of the sample, but increases towards the windings.
The cf peak height is regulated by the azimuthal vortex-free flow,
\textit{i.e.} the height depends on the velocity $v(\Omega_{\rm
f}, R,N)$ at the cylindrical wall. This doubles the frequency of
an azimuthally precessing asymmetry in the cf velocity
$v(\Omega_{\rm f}, R,N)$. For simplicity we assume that $
\varepsilon = {1 \over 2}$ which then compensates for the
frequency doubling owing to the sensitivity pattern.

\begin{figure}[t]
\begin{center}
\centerline{\includegraphics[width=0.9\linewidth]{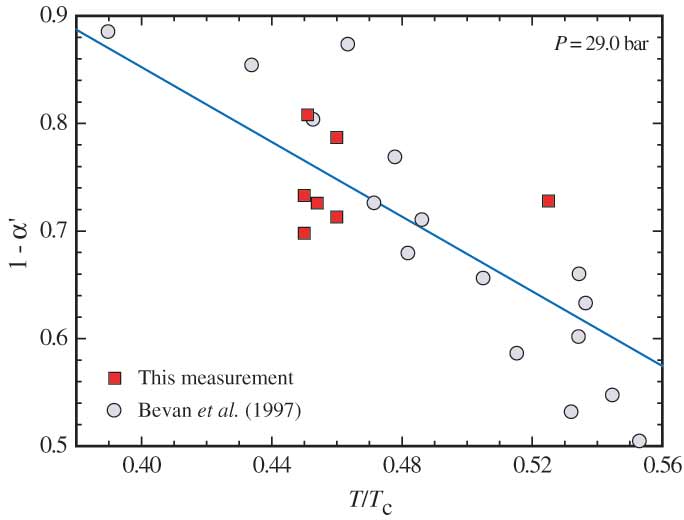}}
\caption{Consistency test on the analysis of the cf peak height
oscillations from seven different measuring runs. The reactive
mutual friction coefficient $1-\alpha^{\prime}$ has been extracted
using Eq.~(\ref{PrecessionFrequency}) and is compared to the
measurements in Ref.~\cite{Bevan}. The line is a fit from
Ref.~\cite{Bevan}.} \label{AlphaPrime}
\end{center}
\vspace{-8mm}
\end{figure}

If we take $\dot{N} \approx 1.2\,$vortices/s and $N = 0$ at the
moment when vortex formation starts (at $t = 170\,$s) in
Fig.~\ref{OscCF_Peak}, we obtain the line through the data in
Fig.~\ref{OscPeriod}. The best fit would have a twice larger
slope, but in view of the large scatter in the signal, which is
only quasi-coherent, the agreement with the calculated line
appears acceptable. Using this approach, we have extracted $1 -
\alpha^{\prime}$ from seven runs with slow oscillatory cf peak
decay. In these seven cases the single-vortex instability proceeds
at a rate $\dot{N} \sim 0.4$
--- 2\,vortices/s, while $\Omega_{\rm f}$ is in the range 0.8 ---
1.2\,rad/s. As seen in Fig.~\ref{AlphaPrime}, the results agree
with the measurements on $1 - \alpha^{\prime} (T)$ in
Ref.~\cite{Bevan}.

\textbf{Discussion:} The oscillatory response in the cf peak
height in Fig.~\ref{OscCF_Peak} is a rare event where several
preconditions seem to be fulfilled: \\ \indent 1) New vortices
have to be created at a slow rate $\dot{N} \sim 1$, which requires
that $ T \approx T_{\rm on}$. \\ \indent 2) In all examples the
longitudinal vortical transit time through the 9\,mm long detector
coil pair is roughly $10\,$s and equal to the oscillation period
in the cf peak height. \\ \indent 3) The oscillations start
immediately when the slow vortex generation switches on. In
Figs.~\ref{FastCF_PeakResponse} and \ref{OscCF_Peak} the
resolution in the measurement of $N$ is roughly one vortex,
\textit{i.e.} the reduction in cf peak height per one vortex is
approximately equal to the amplifier noise (when $N \ll N_{\rm
eq}$). This signal/noise resolution applies to the reduction in cf
peak height, when one new vortex is placed in the center of the
cylinder. Presumably the signal
from a single vortex spiralling along the cylinder wall is larger. \\
\indent 4) To explain the quasi-coherent oscillation, the spiral
motion of several vortices has to be reasonably coherent: the 13
oscillations in Fig.~\ref{OscCF_Peak} cannot be attributed to a
single vortex since the transit time for a vortex end to pass
through the detector coil pair is equal to one period. Three
features may help to create and preserve coherence in spiral
motion: (i) When a vortex end on the cylindrical wall creates a
new loop, one end of the new loop starts spiralling close to the
original end. (ii) Both detectors are close to one of the end
plates of the cylinder and, assuming that the single-vortex
instability occurs randomly everywhere on the cylindrical wall
\cite{de Graaf:2007}, almost all vortices approach the detector
from the direction of the far end of the cylinder. (iii) In
numerical calculations the evolving vortices, which start to
spiral as a close bundle, tend to remain in a bunch while
propagating along the cylinder.

A complete explanation has not been worked out for the mechanisms
which gives rise to the oscillating signal in the cf peak height.
To understand the coherence over many periods of spiral motion we
need better agreement between vortex-dynamics calculations and the
experiment, in particular concerning the generation of new
vortices by the single-vortex instability in a rotating circular
cylinder \cite{de Graaf:2007}. As to the amplitude of the
oscillations, spiral vortex motion gives rise to oscillatory
displacements of the order parameter texture from cylindrical
symmetry, as seen in Fig.~\ref{PrecessVorMotion}. To explain the
signal from such oscillations, numerical calculations are needed
on distortions of flare-out textures from cylindrical symmetry and
on the resulting NMR line shapes.

\textbf{Conclusions:} We have studied the single-vortex
instability of evolving vortices in the turbulent temperature
regime of $^3$He-B. In applied flow the single-vortex instability
generates new vortices and becomes the precursor mechanism which
starts turbulence. In a rotating circular cylinder the instability
occurs while remanent vortices, for instance, expand in spiral
motion towards their stable state as rectilinear vortex lines. The
spiral motion has been examined here with numerical calculations,
but we also argued that oscillations in the NMR response bear
direct evidence for the precessing motion. Other kinds of
oscillating responses from precessing vortex motion have been
reported before. A remarkable example is the unwinding of trapped
circulation around a thin wire suspended along the axis of a
cylindrical container, known as the Vinen vibrating wire
experiment \cite{Zieve}. The spiral motion of an evolving vortex
around a central cluster of vortex lines provides another example
of precessing vortex signals.

{\bf Acknowledgements:} This work was supported by the Academy of
Finland (grants 213496, 124616, 114887), by ULTI research visits
(EU Transnational Access Programme FP6, contract
RITA-CT-2003-505313), and the ESF research program COSLAB.

{\bf Keywords:} quantized vortex, vortex formation, vortex
dynamics, vortex instability, mutual friction, transition to
turbulence, onset temperature of turbulence, precursor of turbulence\\


\vspace{-6mm}

\begin{thebibliography}{9}

\bibitem{Vinen} W.F. Vinen, J. Low Temp. Phys. {\bf 145}, 7 (2007).

\bibitem{Schwarz} K.W. Schwarz, Physica B {\bf 197}, 324
(1994).

\bibitem{NeutronInjection} A.P. Finne, S. Boldarev, V.B. Eltsov, and M. Krusius, J. Low Temp.
Phys. {\bf 135}, 479 (2004).

\bibitem{Precursor} A.P. Finne, V.B. Eltsov, R. H\"anninen, J. Kopu, M. Krusius,
E.V. Thuneberg, and M. Tsubota, Phys. Rev. Lett. {\bf 96}, 85301
(2006).

\bibitem{de Graaf:2007} R. de Graaf, R. H\"anninen, T.V. Chagovets,
V.B. Eltsov, M. Krusius, and R.E. Solntsev, J. Low Temp. Phys.
December (2008); preprint arXiv:0708.3003v2.

\bibitem{ROP} A.P. Finne, V.B. Eltsov, R. H\"anninen, N.B. Kopnin, J. Kopu, M. Krusius,
M. Tsubota, and G.E. Volovik, Rep. Prog. Phys. {\bf 69}, 3157
(2006).

\bibitem{PLTP} V.B. Eltsov, R. de Graaf, R. H\"anninen, M. Krusius,
R.E. Solntsev, V.S. L'vov, A.I. Golov, P.M. Walmsley, Prog. Low
Temp. Phys. Vol XVI, ed. M. Tsubota (Elsevier B.V., Amsterdam,
December 2008); preprint -- arXiv:0803.3225v2.

\bibitem{Bevan} T.D.C. Bevan, A.J. Manninen, J.B. Cook, A.J. Armstrong, J.R. Hook, and
H.E. Hall, J. Low Temp. Phys. {\bf 109}, 423 (1997); Phys. Rev.
Lett. {\bf 74}, 750 (1995).

\bibitem{misalignment} V.M. Ruutu, J.J. Ruohio, M. Krusius, B.
Pla\c{c}ais, and E.B. Sonin, Physica B {\bf 255}, 27 (1998).

\bibitem{Donnelly} R.J. Donnelly, {\it Quantized Vortices in
Helium II} (Cambridge Univ. Press, Cambridge, UK, 1991).

\bibitem{FlightTime} V.B. Eltsov, A.I. Golov, R. de Graaf, R. H\"anninen, M. Krusius,
V. L'vov, and R.E. Solntsev, Phys. Rev. Lett. \textbf{99}, 265301
(2007); A.P. Finne, V.B. Eltsov, R. Blaauwgeers, Z. Janu, M.
Krusius, and L. Skrbek, J. Low Temp. Phys. {\bf 134}, 375 (2004).

\bibitem{TwistedPropagation} V.B. Eltsov, A.P. Finne,  R. H\"anninen, J. Kopu, M.
Krusius, M. Tsubota, and  E.V. Thuneberg, Phys. Rev. Lett. {\bf
96}, 215302 (2006); J. Low Temp. Phys. \textbf{150}, 373 (2008).

\bibitem{movies} See movies at
http://ltl.tkk.fi/research/theory/twist.html

\bibitem{Remnants} R.E. Solntsev, R. de Graaf, V.B. Eltsov, R. H\"anninen, and M. Krusius, J. Low Temp. Phys.
{\bf 148}, 311 (2007).

\bibitem{ExpSetup} A.P. Finne, S. Boldarev, V.B. Eltsov, and M. Krusius, J. Low Temp.
Phys. {\bf 136}, 249 (2004).

\bibitem{Kopu} J. Kopu, R. Schanen, R. Blaauwgeers, V.B. Eltsov, M. Krusius, J.J. Ruohio,
and E.V. Thuneberg, J. Low Temp. Phys. {\bf 120}, 213 (2000).

\bibitem{Ruutu} V.M.H. Ruutu, \"U. Parts, J.H. Koivuniemi, N.B. Kopnin, and M. Krusius,
J. Low Temp. Phys. {\bf 107}, 93 (1997); Europhys. Lett. {\bf 31},
449 (1995).

\bibitem{Kopu-2} J. Kopu, J. Low Temp. Phys. \textbf{146}, 47 (2007).

\bibitem{Zieve} R.J. Zieve, Yu.M. Mukharky, J.D. Close, J.C. Davis, and R.E. Packard, J. Low Temp. Phys. {\bf
91}, 315 (1993); \textit{ibid.} {\bf 90}, 243 (1993).

\end{thebibliography}
\end{document}